\pdfoutput=1
\documentclass[11pt,twoside,a4paper,cmspaper,final,collab]{cms-tdr}
\def\svnVersion{694e5a4}\def\svnDate{2022/05/09}\def\cmsCernNoTag{CERN-EP-2021-244}\def\cmsCernDate{\today}\def\cmsMessage{Published in the Journal of High Energy Physics as \href{http://dx.doi.org/10.1007/JHEP04(2022)144}{\doi{10.1007/JHEP04(2022)144}.}}
\begin{document}\cmsNoteHeader{TOP-20-004}

\cmsNoteHeader{TOP-20-004} 
\title{Measurement of the inclusive \texorpdfstring{\ttbar}{ttbar} production cross section in proton-proton collisions at \texorpdfstring{$\sqrt{s} = 5.02\TeV$}{sqrt(s) = 5.02 TeV} }

\date{\today}

\newcommand{\pb}{\ensuremath{\unit{pb}}\xspace}
\newcommand{\stt}{\ensuremath{\sigma_\ttbar}\xspace}
\newcommand{\pp}{\ensuremath{\Pp\Pp}\xspace}
\newcommand{\mt}{\ensuremath{m_{\cPqt}}\xspace}
\newcommand{\tW}{\ensuremath{\cPqt\PW}\xspace}
\newcommand{\VV}{\ensuremath{\PV\PV}\xspace}
\newcommand{\asmz}{\ensuremath{\alpS(m_{\PZ})}\xspace}
\newcommand{\mur}{\ensuremath{\mu_\mathrm{R}}\xspace}
\newcommand{\muf}{\ensuremath{\mu_\mathrm{F}}\xspace}
\newcommand{\abseta}{\ensuremath{\abs{\eta}}\xspace}
\newcommand{\hdamp}{\ensuremath{h_\text{damp}}\xspace}
\newcommand{\Irel}{\ensuremath{I_{\text{rel}}}\xspace}

\newlength\cmsTabSkip\setlength{\cmsTabSkip}{1ex}
\providecommand{\cmsTable}[1]{\resizebox{\textwidth}{!}{#1}}

\abstract{
The top quark pair production cross section is measured in proton-proton collisions at a center-of-mass energy of 5.02\TeV. 
The data were collected in a special LHC low-energy and low-intensity run in 2017, and correspond to an integrated luminosity of 302\pbinv.
The measurement is performed using events with one electron and one muon of opposite charge, and at least two jets.
The measured cross section is $60.7 \pm 5.0 \stat \pm 2.8 \syst \pm 1.1 \lum \pb$.
A combination with the result in the single lepton + jets channel, based on data collected in 2015 at the same center-of-mass energy and corresponding to an integrated luminosity of 27.4\pbinv, is then performed. The resulting measured value is $63.0 \pm 4.1 \stat \pm 3.0\, \text{(syst+lumi)} \pb$, in agreement with the standard model prediction of $66.8^{+2.9}_{-3.1}\pb$. 
}

\hypersetup{
pdfauthor={CMS Collaboration},%
pdftitle={Measurement of the inclusive ttbar production cross section in proton-proton collisions at sqrt(s) = 5.02 TeV},%
pdfsubject={CMS},
pdfkeywords={CMS,  top, top cross section}}

\maketitle 

\section{Introduction} \label{sec:introduction}
The top quark is the most massive elementary particle in the standard model (SM). The study of its production and properties is one of
the core elements of the CERN LHC physics program. At the LHC, top quarks are primarily produced in pairs
(\ttbar), and the \ttbar production cross section is sensitive to the gluon parton distribution function
(PDF) of the proton~\cite{Sirunyan:2017azo} and to the top quark pole mass~\cite{Sirunyan:2018goh}. The ATLAS, CMS, and LHCb Collaborations have performed
several cross section measurements with increasing precision in a variety of decay channels at four proton-proton (\pp) collision
energies~\cite{Sirunyan:2017ule,ATLAStt12,Aad:2014kva,Khachatryan:2016mqs,Khachatryan:2016yzq,ATLAS:2019hau,Khachatryan:2015uqb,Khachatryan:2016kzg,CMS:2021vhb,Aaij:2018imy},
as well as in proton-nucleus~\cite{Sirunyan:2017xku} and nucleus-nucleus~\cite{Sirunyan:2020kov} collisions.

The first measurement of the \ttbar production cross section, \stt, in \pp collisions at a center-of-mass energy of 5.02\TeV, was performed by the CMS experiment analyzing events with one or two leptons ($\ell=$ electron or muon) and at least two jets, using a data sample taken in 2015 that corresponds to an integrated luminosity of 27.4\pbinv.
The measurement of $\stt=69.5 \pm 6.1 \stat \pm 5.6 \syst \pm 1.6 \lum\pb$ was obtained from the combination of the results in the dilepton and single lepton decay channels~\cite{Sirunyan:2017ule}.  

During the year 2017, the LHC delivered a subset of \pp collisions at $\sqrt{s}=5.02\TeV$ and CMS collected a data sample
corresponding to 302\pbinv, an increase in integrated luminosity of more than an order of magnitude compared to the data set recorded in 2015. 
A distinct feature of this data sample is the low number of additional interactions per bunch crossing (pileup) with respect to the standard operating conditions of the LHC. 
We present here a measurement of \stt using events with two opposite-charge different-flavor leptons, \ie, one electron
and one muon ($\Pepm{}\Pgm^\mp$), and at least two jets. The cross section is extracted using a counting experiment and the result is then combined with the measurement in the $\ell$+jets final state contained in Ref.~\cite{Sirunyan:2017ule}.

This paper is organized as follows. A brief description of the CMS detector, and of the Monte Carlo (MC) simulation samples, are given in Section~\ref{sec:CMS},
followed by the object and event selection in Section~\ref{sec:EventSelection}. The background estimation methods are covered in
Section~\ref{sec:Background} and the systematic uncertainties in Section~\ref{sec:Systematics}. Results are discussed
in Section~\ref{sec:Results} and the summary is given in Section~\ref{sec:Summary}. Tabulated results are provided in HEPData~\cite{hepdata}.

\section{The CMS detector and Monte Carlo simulation} \label{sec:CMS}
The central feature of the CMS apparatus is a superconducting solenoid of 6\unit{m} internal diameter,
providing a magnetic field of 3.8\unit{T}. Within the solenoid volume are a silicon pixel and strip tracker,
a lead tungstate crystal electromagnetic calorimeter (ECAL), and a brass and scintillator hadron calorimeter (HCAL), 
each composed of a barrel and two endcap sections. Forward calorimeters extend the pseudorapidity ($\eta$) coverage
provided by the barrel and endcap detectors. Muons are detected in gas-ionization chambers embedded in the steel
flux-return yoke outside the solenoid. A more detailed description of the CMS detector, together with a definition
of the coordinate system used and the relevant kinematic variables, can be found in Ref.~\cite{Chatrchyan:2008zzk}.

Simulated event samples are used to define the analysis strategy, to estimate the background contribution, and to evaluate
efficiencies and uncertainties. The samples used in the analysis are summarized in Table~\ref{tab:mc}. The propagation
of the generated particles through the CMS detector and the modeling of the detector response are performed using
\GEANTfour~\cite{Agostinelli:2002hh}.

Simulated \ttbar events are generated at next-to-leading order (NLO) in quantum chromodynamics (QCD) using \POWHEG (v2)~\cite{Frixione:2007vw,Alioli:2010xd,Frixione:2007nw},
assuming a top quark mass \mt of 172.5\GeV. The events are then interfaced with {\PYTHIA8} (v230)~\cite{Sjostrand:2014zea}
with the ``CP5'' tune~\cite{Sirunyan:2019dfx} for parton showering, hadronization, and the
underlying event description. For the study of the acceptance dependence on \mt, alternative generator-level samples have
been used with $\mt=166.5$ and 178.5\GeV. The NNPDF3.1~\cite{Ball:2017nwa} next-to-next-to-leading-order (NNLO) PDFs are used. A similar setup is used for the simulation of the single top quark production in association with a \PW boson (\PQt{}\PW).

The \MGvATNLO (v2.4.2) generator~\cite{Alwall:2014hca}, interfaced with {\PYTHIA8} for parton showering, is used to simulate \PW boson production with additional jets ({\PW}+jets), and Drell--Yan (DY) quark-antiquark annihilation
into lepton-antilepton pairs through \PZ boson or virtual-photon exchange.
The simulation is performed at NLO in QCD and includes up to two extra partons at the matrix element (ME) level. The FxFx matching scheme~\cite{Frederix:2012ps} is used to merge jets from the ME calculations and the parton shower (PS). Diboson (\VV, with \PV = \PW or \PZ) events are simulated at NLO in QCD with \POWHEG (v2). When available, higher-order cross sections are used instead of those of the generator, as shown in Table~\ref{tab:mc}.

\begin{table}[htp]
  \topcaption{
    Summary of MC samples used to model the signal and background processes. The column ``Cross 
    section  order'' corresponds to the QCD or electroweak (EW) precision used to normalize the distributions provided by the
    generators. Where no reference is given, the precision of the MC simulation is kept.}
  \label{tab:mc}
  \centering
    \begin{tabular}{ccl}
      Process        & Generator + parton shower &  Cross section order  \\
      \hline 
      \ttbar & \POWHEG+ \PYTHIA8 & NNLO+NNLL~\cite{Czakon:2011xx,Czakon:2013goa} \\
      \tW &  \POWHEG+ \PYTHIA8  &  Approximate NNLO~\cite{Kidonakis:2015nna}\\ 
      {\PW}+jets &  \MGvATNLO+ \PYTHIA8 & NNLO[QCD]+NLO[EW]~\cite{Melnikov:2006kv} \\ 
      DY  &  \MGvATNLO+ \PYTHIA8  & NNLO[QCD]+NLO[EW]~\cite{Melnikov:2006kv} \\ 
      \VV  &  \POWHEG+ \PYTHIA8 &  NLO \\ 

    \end{tabular}
\end{table}

The SM prediction for \stt at 5.02\TeV is $66.8^{+1.9}_{-2.3}\,\text{(scale)} \pm 1.7\,\text{(PDF)} ^{+1.4}_{-1.3}\,(\asmz)\,\pb$ for $\mt= 172.5\GeV$ and a strong coupling at the \PZ boson mass, \asmz, of $0.118 \pm 0.001$~\cite{PDG2020}. This prediction is calculated with the \TOPpp program~\cite{Czakon:2011xx} at NNLO in perturbative QCD including soft-gluon resummation at next-to-next-to-leading-log (NNLL) approximation~\cite{Czakon:2013goa}. The first uncertainty reflects variations in the factorization (\muf) and renormalization (\mur) scales. The second and third uncertainties are associated with possible choices of PDFs and the \alpS value respectively, using the NNPDF3.1~\cite{Ball:2017nwa} NNLO PDF sets that include top quark measurements. The expected integrated event yields for signal in all figures and tables are normalized to the predicted cross section.

The simulated samples include multiple \pp collisions occurring in the same bunch crossing (pileup), with a distribution that matches that observed in data, with an average of about two pileup collisions per bunch crossing.

\section{Object reconstruction and event selection}  \label{sec:EventSelection}
Events of interest are selected online using a two-tiered trigger system~\cite{Khachatryan:2016bia,Sirunyan:2020zal}. The first level (L1),
composed of custom hardware processors, uses information from the calorimeters and muon detectors to select events
at a rate of around 100\unit{kHz} within a fixed latency of less than 4\unit{\mus}. The second level, known as the high-level
trigger, consists of a farm of processors running a version of the full event reconstruction software optimized
for fast processing, and reduces the event rate to around 1\unit{kHz} before data storage. Only events that fired at
least one of the single-lepton triggers with transverse momentum (\pt) thresholds greater than 17 (12)\GeV in the case of electrons (muons) are considered.

Events may contain multiple primary vertices, corresponding to pileup collisions. The candidate vertex with the largest value of summed physics-object $\pt^2$ is taken to be the primary \pp interaction vertex. The physics objects are the jets, clustered using the jet finding algorithm~\cite{Cacciari:2008gp,Cacciari:2011ma} using tracks assigned to candidate vertices as inputs, and the associated missing transverse momentum, taken as the negative vector sum of the \pt of those jets.

The particle-flow algorithm~\cite{CMS-PRF-14-001} aims to reconstruct and identify each
individual particle in an event, with an optimized combination of information from the
various elements of the CMS detector. The energy of electrons is determined from a
combination of the electron momentum as measured by the tracker, the energy of the corresponding ECAL cluster, and the energy sum of
all bremsstrahlung photons spatially compatible with originating from the electron track.
The energy of muons is obtained from the curvature of the corresponding track. The energy
of charged hadrons is determined from a combination of their momentum measured in the
tracker and the matching ECAL and HCAL energy deposits, corrected for the response function
of the calorimeters to hadronic showers. Finally, the energy of neutral hadrons is obtained
from the corresponding corrected ECAL and HCAL energies.

Jets are clustered from these reconstructed particles using the anti-\kt algorithm~\cite{Cacciari:2008gp, Cacciari:2011ma} with a distance parameter of 0.4. The jet momentum is determined as the vectorial sum of all particle momenta in the jet, and is found from simulation to be, on average, within 5 to 10\% of the true momentum over the whole \pt spectrum and detector acceptance.
Jet energy corrections are derived from simulation studies
so that the average measured energy of jets becomes identical to that of particle-level
jets. Measurements of the momentum balance are used to determine any residual
differences between the jet energy scale in data and in simulation, and appropriate
corrections are made~\cite{Khachatryan:2016kdb}. These corrections were derived using the full low-pileup data set of \pp collisions at 5.02\TeV. Additional selection criteria are applied to remove jets potentially dominated by instrumental effects or reconstruction failures~\cite{CMS:2017wyc}.

Electron candidates are required to satisfy $\abseta<2.5$ and $\pt>10\GeV$. To identify electrons, requirements are
placed on a multivariate discriminant based on the shower shape and track
quality of the electron candidates~\cite{CMS:2020uim}. Electron candidates that are matched to a secondary vertex
consistent with a photon conversion, or have a missing hit in the inner layer of the tracker are vetoed.

Reconstructed muon candidates are required to have $\abseta<2.4$ and $\pt>10\GeV$, and must fulfill criteria on the
geometrical matching between the tracks reconstructed by the silicon tracker and the muon system, and on the quality of the global
fit~\cite{Sirunyan:2018fpa}.

Lepton candidates must be consistent with originating from the primary vertex which is ensured by requiring that the
transverse (longitudinal) impact parameter should not exceed 0.05 (0.10)\unit{cm}. Furthermore, the significance
of the three-dimensional impact parameter must be smaller than 8. Electrons and muons must also satisfy
a requirement on their relative isolation (\Irel), defined as the scalar \pt sum of all the particles inside a cone around the lepton direction, excluding the lepton itself, divided by the lepton \pt.
The cone size, defined as $\Delta R=\sqrt{\smash[b]{(\Delta\eta)^2+(\Delta\phi)^2}}$, where $\phi$ is the azimuthal angle, changes as a function of the lepton \pt as $\Delta R(\pt) = 10\GeV /\pt$ if $50\GeV < \pt < 200\GeV$, $\Delta R = 0.2$ if $\pt\leq 50\GeV$ and $\Delta R = 0.05$ otherwise.
Electrons (muons) must satisfy the condition $\Irel < 0.085\, (0.325)$. 

To reject leptons originating from hadron decays, or misidentified leptons, also referred to as
``nonprompt'' leptons, from those produced in the decay of the electroweak bosons (``prompt''), a
gradient boosted decision tree (BDT) is used, trained using MC simulation, to distinguish prompt from nonprompt leptons~\cite{Sirunyan:2020icl}.
 This BDT uses the properties of the jet containing the lepton, as returned by the jet clustering algorithm: its
\PQb tagging score, the ratio of the lepton \pt to that of the jet, and the momentum of the jet transverse to the
lepton direction. Other input variables are the lepton \pt, $\eta$, \Irel, longitudinal and transverse
impact parameters, and the significance of the three-dimensional impact parameter. 
In addition, the previously mentioned multivariate discriminant for electrons and the muon segment compatibility for muons are used as input variables~\cite{Sirunyan:2018fpa}.
To further suppress nonprompt leptons originating from b quark decays, leptons associated with a jet 
satisfying the loose working point of the DeepCSV \PQb tagging algorithm~\cite{Sirunyan:2017ezt} are rejected. 

The \ttbar candidate events are required to have at least two leptons (one electron and one muon) with opposite charge and at
least two jets. Only jets with $\pt>25\GeV$, $\abseta<2.4$, and containing no selected leptons are considered.
To ensure efficient triggering of the events, the leading lepton is required to have $\pt>20\GeV$. In addition, events
must have a dilepton invariant mass above 20\GeV to reduce the background from photon conversions and low-mass resonances.

\section{Background estimation} \label{sec:Background}

Background events arise mainly from \tW, DY, and \VV production in which at least two prompt leptons
emerge from the \PZ or \PW boson decays. The \tW and \VV contributions are estimated from simulation.

The DY event yield is estimated from data using the $R_{\text{out}/\text{in}}$ method~\cite{Khachatryan:2016mqs}, 
where events with same-flavor leptons are used to normalize the yield of different-flavor pairs from DY production of \Pgt lepton pairs. A
data-to-simulation normalization factor is estimated from the number of events in data within a 15\GeV window around the \PZ boson mass
and extrapolated to the number of events outside the \PZ boson mass window with corrections applied using control regions enriched in DY events
in data. This factor is measured to be $0.91 \pm 0.01$. The stability of the
method against a potential mismodeling of the jet multiplicity is checked and found to be within 30\%, which will be considered as an
extra systematic uncertainty in this background estimation.

Other residual background sources, such as \ttbar where only one of the \PW bosons decay leptonically
or {\PW}+jets events, may contaminate the signal sample when a jet is misreconstructed
as a lepton, or contains a lepton from a \cPqb{}/\cPqc{} hadron decay, incorrectly identified as a prompt lepton. 
These events are grouped into the nonprompt lepton category, together with meson decays and photon conversions; their contribution is estimated with simulated \ttbar events with at least one \PW boson decaying into jets and {\PW}+jets events.

\begin{figure}[htb!]
  \centering
  \includegraphics[width=.49\textwidth]{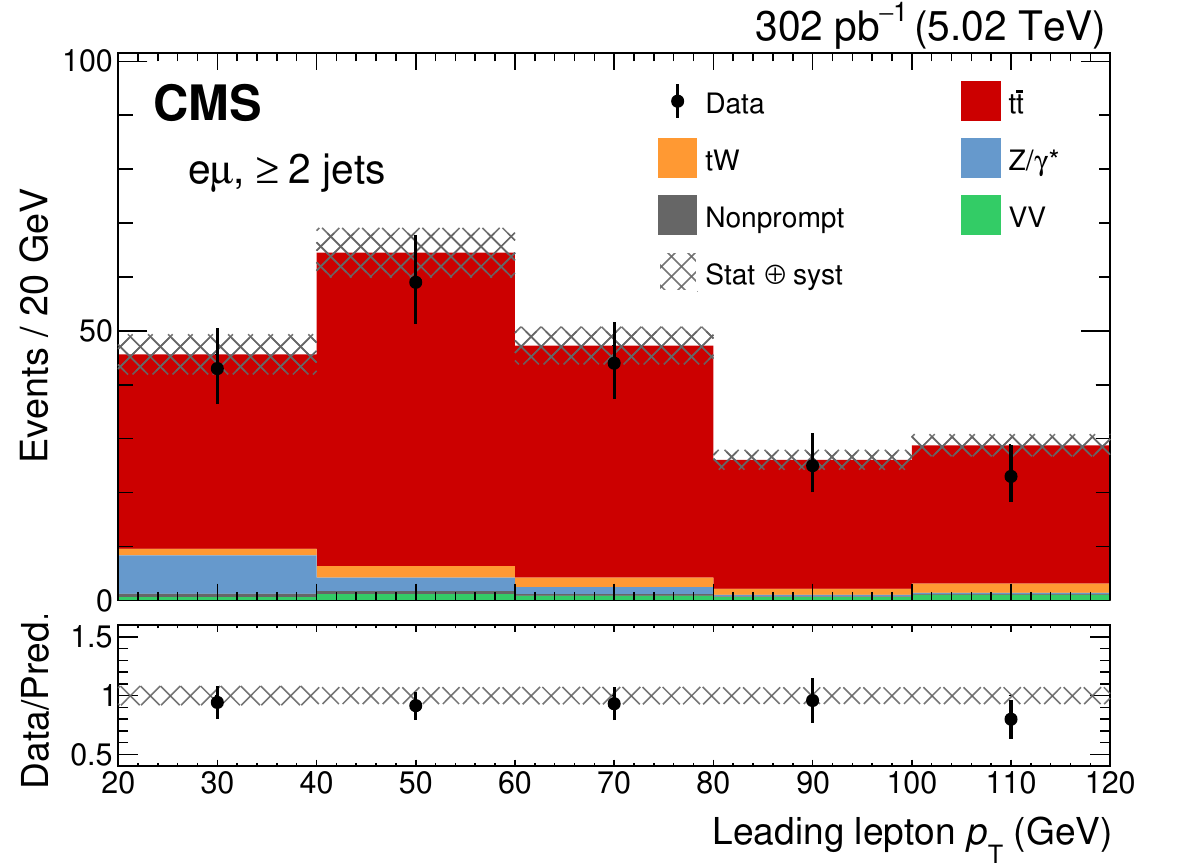}
  \includegraphics[width=.49\textwidth]{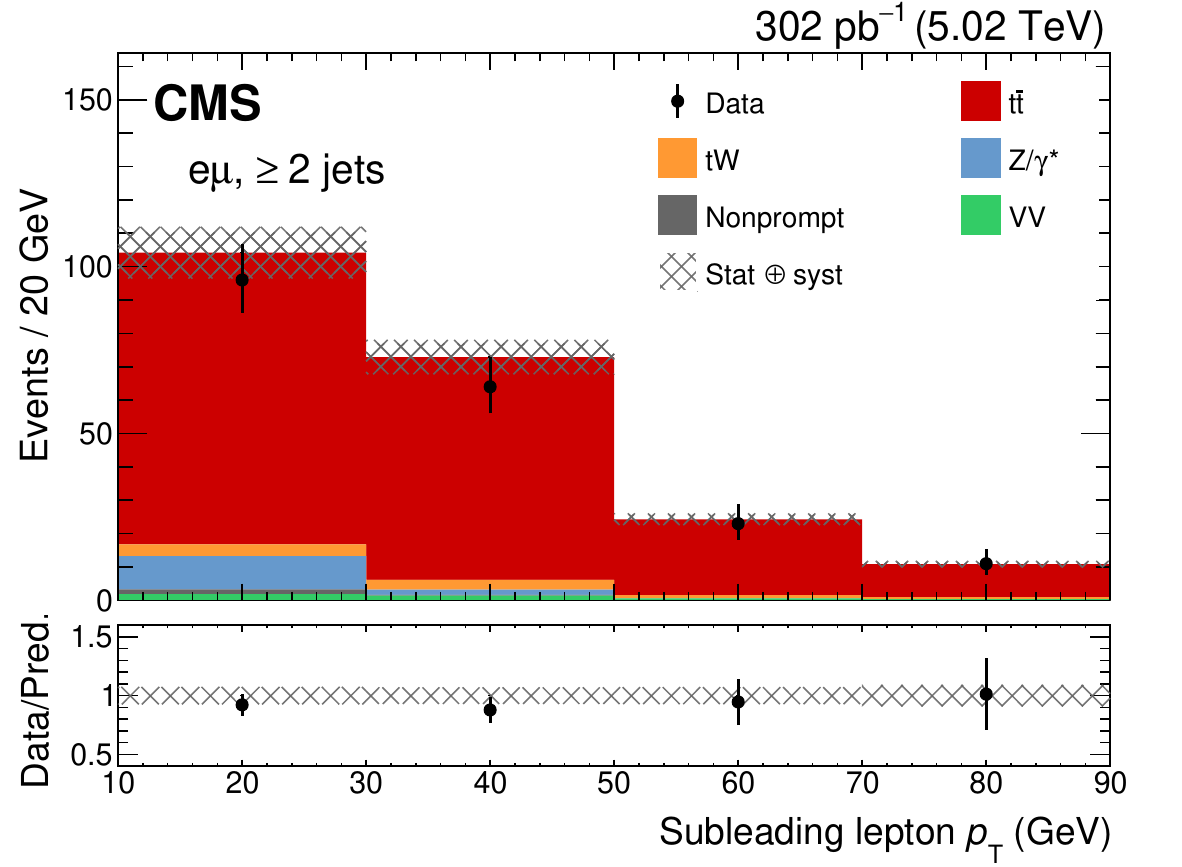} \\
  \includegraphics[width=.49\textwidth]{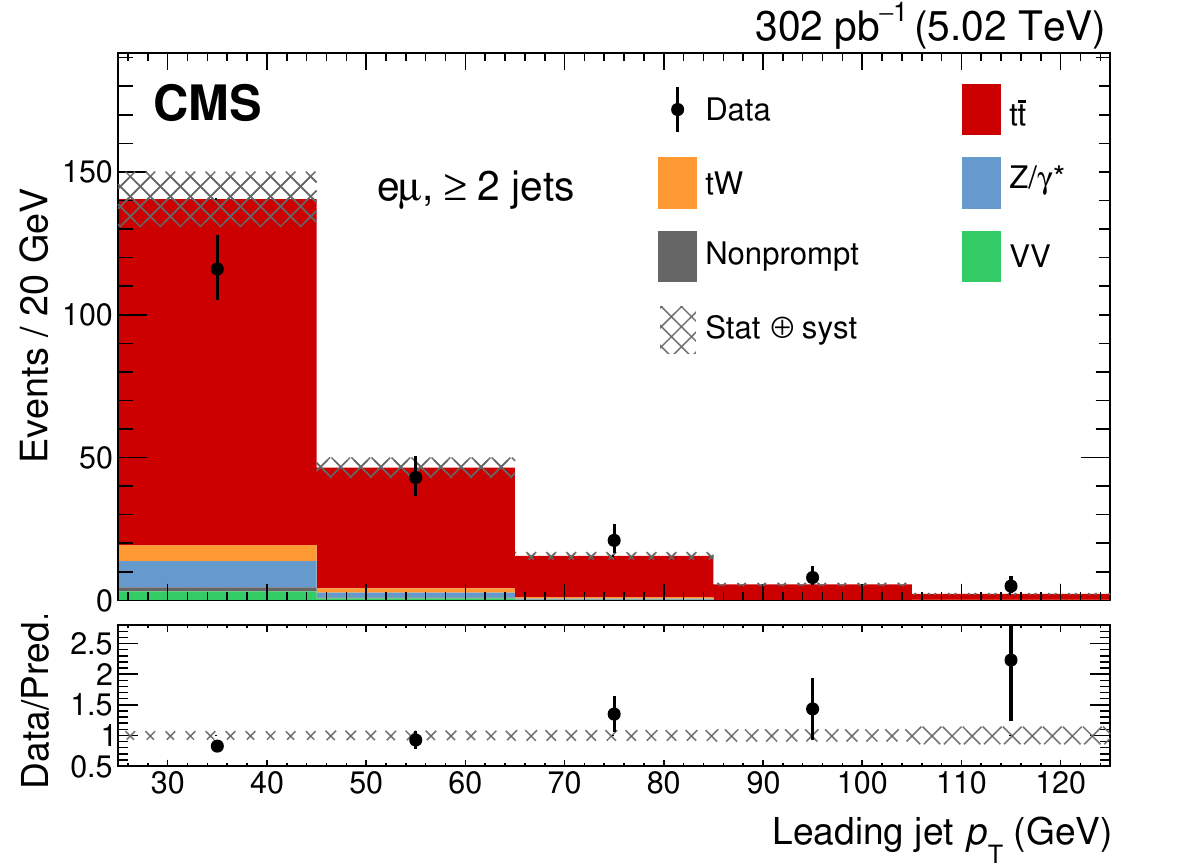} 
  \includegraphics[width=.49\textwidth]{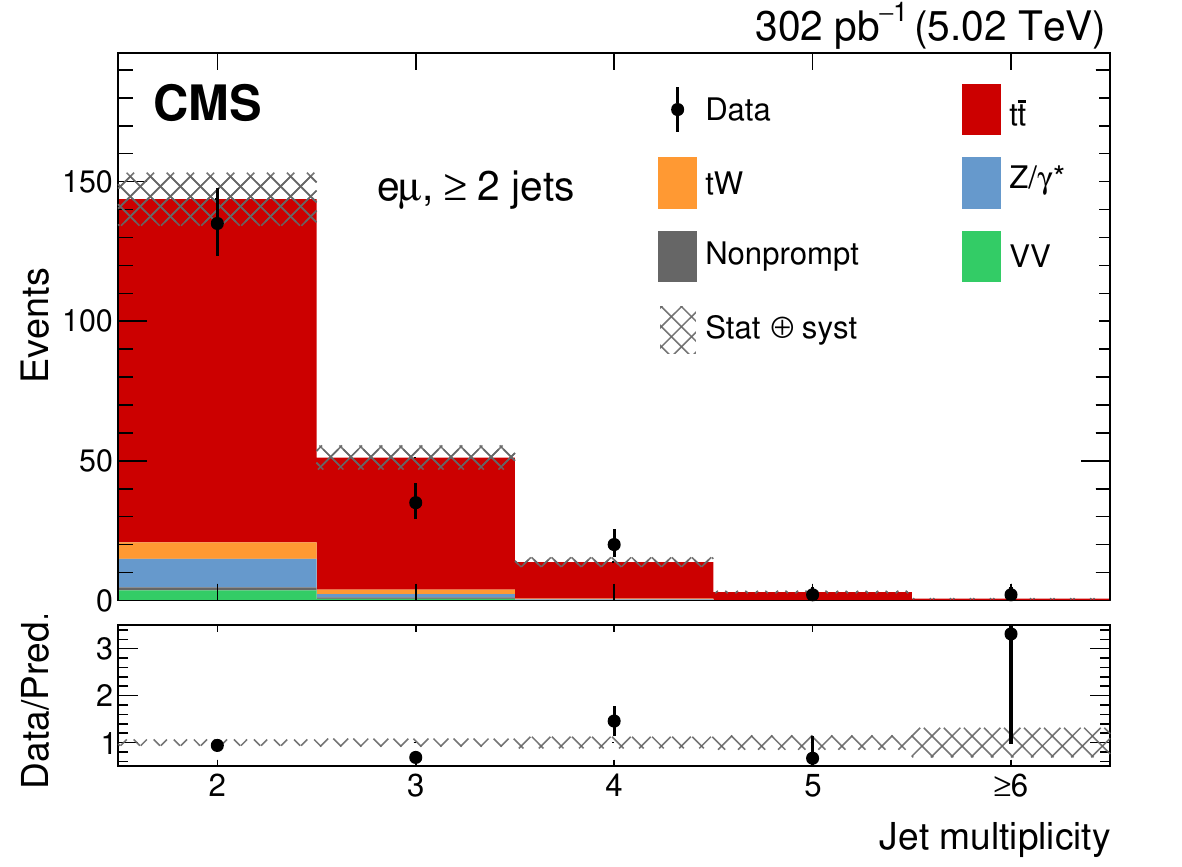}
\caption{Leading lepton \pt (upper left), sub-leading lepton \pt (upper right), leading jet \pt (lower left), and jet multiplicity (lower right) in the selected events. The hatched band corresponds to systematic and statistical uncertainties summed in quadrature. The lower panels show the data-to-prediction ratio. The last bin in each distribution includes overflow events.}
\label{fig:events}
\end{figure}

Figure~\ref{fig:events} shows the \pt of the two leptons and of the leading jet, and the jet multiplicity of the selected
events. The data are compared to the sum of the expected signal and background distributions for the \ttbar signal and individual backgrounds, which are derived either from simulated samples or from data, as described above. The expected distributions describe the data within the experimental uncertainties.

\section{Systematic uncertainties} \label{sec:Systematics}

The measurement of \stt is affected by sources of systematic uncertainty related to detector effects or
theoretical assumptions. Each source of systematic uncertainty is evaluated by repeating the \stt extraction with variations of the input parameters by $\pm$1 standard deviations~(experimental uncertainty) or from dedicated simulation samples with different settings (theoretical uncertainty). The experimental uncertainties affect mostly the efficiency, while modeling uncertainties affect both the acceptance and the efficiency. The total uncertainty is calculated by adding the effects of all the individual systematic components in quadrature, assuming they are independent. The sources of systematic uncertainty are described in detail below:

\begin{description}
\item[Lepton-related uncertainty:]  Lepton reconstruction and identification efficiencies, as well as energy scale and resolution, are
  measured using \PZ boson events in low-pileup data and simulation of \pp collisions at 5.02\TeV~\cite{CMS:2020uim,Sirunyan:2018fpa}. Correction factors are then applied to the simulation to improve the agreement
  with the data.  The uncertainty in these corrections is propagated to the \stt measurement. For the selected events, the
  trigger efficiency is very close to 1, as the trigger conditions are satisfied redundantly by both leptons in most of the events. The deviation of the efficiency from unity, obtained from a \ttbar simulated sample, is used as the associated uncertainty in \stt.

\item[Jet-related uncertainty:] The impact of the uncertainty in jet energy scale (JES) is estimated from the change in the number of simulated \ttbar events selected after changing the jet momenta within the JES uncertainties~\cite{Khachatryan:2016kdb}. The effect of the jet energy resolution (JER) is determined by an $\eta$-dependent variation of the JER scale factors within their uncertainty.

\item[L1 prefiring:] During the 2017 data taking, a gradual shift in the timing of the inputs of the ECAL L1 trigger in the region $\abseta>2.0$ caused a trigger inefficiency~\cite{Sirunyan:2020zal}. Simulations are corrected to mimic this behavior in data and the uncertainty in these corrections is propagated to \stt by varying the correction within the associated uncertainty.

\item[Scale choice:] The uncertainty related to the missing higher-order diagrams in \POWHEG is estimated by varying the default \muf and \mur choices independently by a factor of 2 and 1/2. As uncertainty in the signal acceptance is
  assigned the maximum difference of each variation from the nominal values, excluding variations of the scales in opposite directions.

\item[Parton shower scale:] The effect of the choice of PS scale is studied by changing the scale used for the initial-
  and final-state radiation by a factor 2 and 1/2 with respect to its default value. The maximum variation with respect to the central
  sample is taken as the uncertainty.

\item[Matrix element and PS matching (\hdamp):] The impact of the ME and
	PS matching, which is parameterized by the \POWHEG generator as \hdamp, with a nominal value of ${(1.4^{+0.9}_{-0.5}) \mt}$~\cite{Sirunyan:2019dfx},
  is calculated by varying this parameter within the uncertainties, using dedicated samples. The variation with respect to the central value
  of the signal acceptance at particle level is considered as the uncertainty in the \stt extraction. 

\item[Parton distribution functions:] The uncertainty due to the proton PDFs is evaluated by reweighting simulated
  signal events using the replicas of the NNPDF3.1 set~\cite{Ball:2017nwa}.
  The variations consist of a central PDF and 100 replicas, for which the root mean square of all differences of the resulting \stt with respect to the central value is taken as the uncertainty. 
  Two extra variations corresponding to different \asmz choices are added in quadrature~\cite{Butterworth:2015oua}.
  
\item[Underlying event tune:] The parameters of \PYTHIA are adjusted to model the measured underlying event tune~\cite{Sirunyan:2019dfx}. The uncertainty
  is calculated by varying these parameters within their uncertainties in dedicated simulated samples. The variation with respect to the
  central value of the signal acceptance is taken as the uncertainty. 

\item[Background normalization:] The uncertainty in the \tW  and \VV cross sections is taken to be 20 and 30\%, respectively,
  based on the theoretical uncertainties and the effect of finite size of the simulated samples. To the nonprompt background estimation is assigned a 50\% uncertainty to account for
  possible mismodeling of the data in simulation. As explained in Section~\ref{sec:Background}, a 30\% uncertainty is considered for the DY background normalization.

\item[Pileup and integrated luminosity:] The uncertainty assigned to the number of pileup events in simulation is calculated by varying
  the total inelastic \pp cross section by 4.6\%~\cite{Sirunyan:2018nqx}. The impact of this uncertainty on the result is negligible, as the
  number of pileup events is also small. The uncertainty in the measurement of the integrated luminosity is estimated to be
  1.9\%~\cite{bib:LUM-19-001}.

\end{description}

Table~\ref{tab:breakdown_comb} summarizes the sources of systematic and statistical uncertainties in the measured \stt, as obtained using Eq.~\eqref{eq:xsec}, explained in the next section. The result is dominated by the statistical uncertainty, while the uncertainty in the JES and the DY background estimate constitute the largest systematic uncertainties.

\begin{table}[htb!]
\topcaption{Summary of the systematic and statistical relative uncertainties for the inclusive \ttbar cross section measurement.} \label{tab:breakdown_comb}
\centering
\begin{tabular}{l c}
Source         & $\Delta\stt/\stt$ (\%) \\
  \hline
 Electron efficiency    &  1.6            \\ 
 Muon efficiency        &  0.6            \\ 
 Trigger efficiency     &  1.3            \\ 
 JES                    &  2.2           \\ 
 JER                    &  1.2           \\ 
 L1 prefiring           &  1.4            \\ 
 \mur, \muf scales &  0.2            \\ 
 Final-state radiation      &  1.1             \\ 
 Initial-state radiation    &  $<0.1$          \\ 
 \hdamp          &  1.0             \\ 
 PDF$\oplus\asmz$ &  0.3            \\ 
 Underlying event tune      &  0.7            \\ 
            \tW  &  1.0           \\ 
      Nonprompt leptons &  0.4           \\ 
             Drell--Yan &  1.8           \\ 
              \VV  &  0.8           \\ [\cmsTabSkip]
 Total systematic uncertainty &  4.3     \\ [\cmsTabSkip]
 Integrated luminosity      &  1.9       \\  [\cmsTabSkip]
 Statistical uncertainty    &  8.2       \\ 
\end{tabular}
\end{table}

\section{Results} \label{sec:Results}

The \ttbar production cross section is extracted via the expression

\begin{equation}
	\stt=\frac{N-N_\text{bkg}}{\varepsilon\mathcal{A}\cal{B}\mathcal{L}} ,
\label{eq:xsec}
\end{equation}

where $N$ is the number of observed events, $N_\text{bkg}$ is the number of estimated background events, $\mathcal{L}$ is the integrated luminosity, $\mathcal{B}=3.19\%$ is the SM value~\cite{PDG2020} of the branching fraction of a \PW boson pair to $\Pepm{}\Pgm^\mp$, including decays through \Pgt leptons, $\mathcal{A}$ is the total acceptance, defined as the fraction of all generated $\ttbar \to \Pepm{}\Pgm^\mp$ events fulfilling the aforementioned kinematic selection criteria, and $\varepsilon$ is the reconstruction efficiency. The acceptance is estimated from simulation and is found to be $0.54\pm 0.01$. The efficiency is estimated from simulation, after applying all the correction factors for leptons and jets to match the performance of the data, and is measured to be $0.53\pm 0.02$. Table~\ref{tab:yields} shows the total number of events observed in data together with the total number of expected signal and background events.

\begin{table}[hbt]
    \topcaption{Event yields for all the processes at the final level of selection. The uncertainty corresponds to the quadratic sum of the statistical and systematic sources.}
    \label{tab:yields}
    \centering
    \begin{tabular}{@{} l r@{\,$\pm$\,}l }
      Process           & \multicolumn{2}{l}{Event yield} \\
      \hline

      \tW        &   8 & 2  \\
      Nonprompt leptons &   2 & 1  \\
      DY                &  10 & 4  \\
      VV                &   4 & 1  \\ [\cmsTabSkip]
      Total background  &  24 & 4  \\ [\cmsTabSkip]
      \ttbar            & 187 & 9  \\ [\cmsTabSkip]
     Data               &  \multicolumn{2}{l}{194}   \\
    \end{tabular}
\end{table}

The measured inclusive cross section for a top quark mass of 172.5\GeV is

\begin{equation*}
 \stt =  60.7 \pm 5.0 \stat \pm 2.8 \syst \pm 1.1 \lum \pb.
\end{equation*}

The fiducial cross section ($\stt^{\text{fid}}$) is measured for events containing one electron and one muon with $\pt>10\GeV$ and $\abseta< 2.4$, invariant mass of the pair of at least 20\GeV, a leading lepton \pt of at least 20\GeV, and at least two jets with $\pt> 25\GeV$ and $\abseta<2.4$. For the fiducial cross section measurement, an estimate of the uncertainties similar to that shown in Table~\ref{tab:breakdown_comb} is made. The resulting value is $\stt^{\text{fid}} = 1.05 \pm 0.09 \stat \pm 0.05 \syst \pm 0.02 \lum \pb$.

The acceptance has been predicted for $\mt= 166.5$ and 178.5\GeV and is parameterized as a linear function of \mt. The cross section varies by $\mp 0.30 \pb$ when the top quark mass changes by $\pm 0.5\GeV$.

The result is combined with that obtained in the $\ell$+jets decay
channel of Ref.~\cite{Sirunyan:2017ule}, corresponding to an integrated luminosity of 27.4\pbinv. The result obtained in the dilepton decay channel of Ref.~\cite{Sirunyan:2017ule} was not added to the combination, as its contribution would be negligible.
We determine the combined \stt using the best linear unbiased estimator
(BLUE) method~\cite{Lyons:1988rp,Valassi:2013bga}. The 2015 measurement in the $\ell$+jets channel yielded a cross section
of $\stt = 68.9\pb$ with a total uncertainty of 13\%, dominated by the statistical uncertainty. Most sources of experimental uncertainty are considered as uncorrelated, given that the data sets and background estimation methods are different, with the exception of the uncertainties on the \tW background and the scale choice, which are considered as fully correlated. The modeling uncertainties are taken as fully correlated. The resulting cross section is 

\begin{equation*}
\stt^{\text{comb}} =  63.0 \pm 4.1 \stat \pm 3.0 \, \text{(syst+lumi)} \pb, 
\end{equation*}

where the total uncertainty of 8.0\% is the quadrature sum of the individual sources of uncertainty.
The weights of the individual measurements, to be understood in the sense of Ref.~\cite{Valassi:2013bga}, are 27 and 73\% for the $\ell$+jets~\cite{{Sirunyan:2017ule}} and the measurement presented in this paper, respectively. This result is in agreement with the SM prediction.   

The combined result is found to be robust by performing an iterative variant of the BLUE method~\cite{Lista:2014qia} and varying some 
assumptions on the correlations of different combinations of systematic uncertainties. Also, the correlations between the
nuisance parameters in both channels have been checked and found to have a negligible impact.

\begin{figure}[!htb]
\centering
\includegraphics[width=0.98\textwidth]{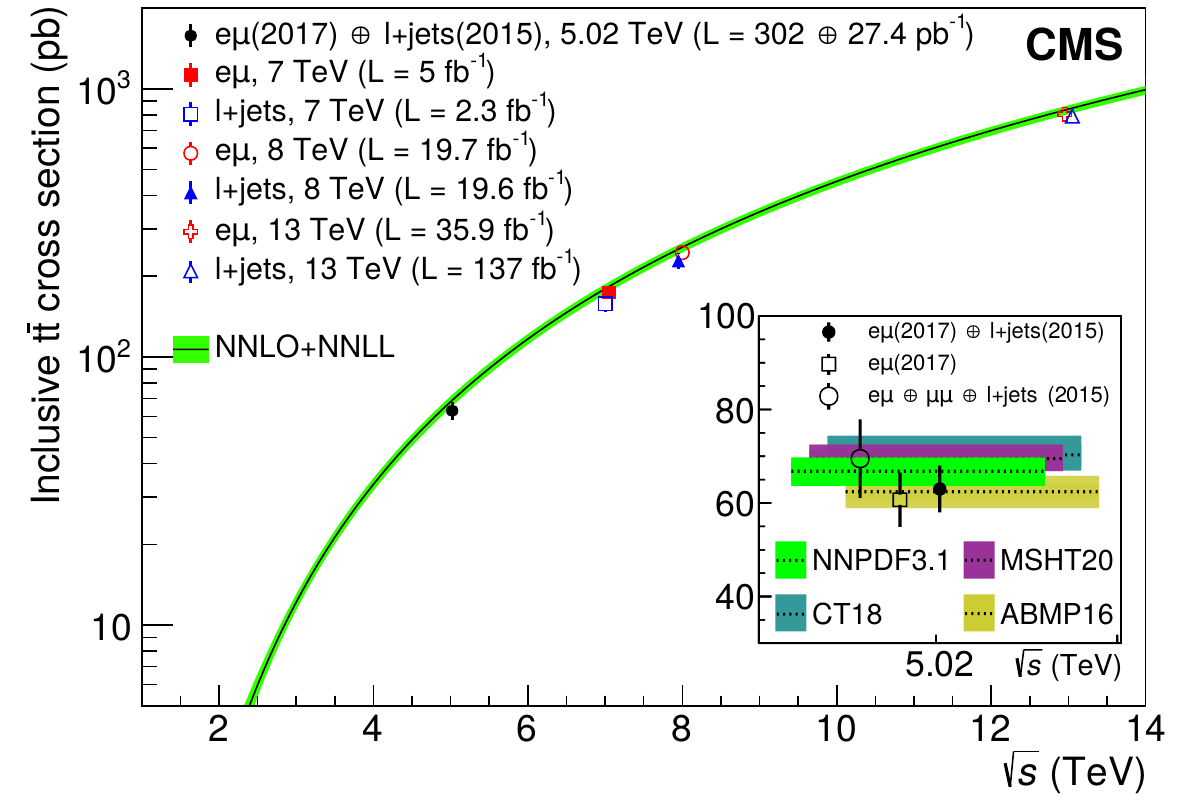}
\caption{
  Inclusive \ttbar cross section in \pp collisions as a  function of the center-of-mass energy 
  in the separate $\ell$+jets and dilepton channels along with the combined measurement at 5.02\TeV presented in this analysis are displayed. 
  Some of the previous CMS measurements at $\sqrt{s} = 7$, 8~\cite{Khachatryan:2016mqs,Khachatryan:2016yzq}, and
  13\TeV~\cite{Sirunyan:2018goh,CMS:2021vhb} are also shown.
  The NNLO+NNLL theoretical prediction~\cite{Czakon:2013goa} using the NNPDF3.1~\cite{Ball:2017nwa} PDF set with $\asmz=0.118$ and $\mt=172.5\GeV$ is shown in the main plot. In the inset, 
  predictions at 5.02\TeV using the MSHT20~\cite{Bailey:2020ooq}, CT18~\cite{Dulat:2015mca}, and ABMP16~\cite{Alekhin:2017kpj} NNLO PDF sets, the latter with $\asmz=0.115$ and $\mt=170.4\GeV$, are compared, along with the NNPDF3.1 NNLO prediction, to the individual and combined results from this analysis. The vertical bars and bands represent the total uncertainties in the data and in the predictions, respectively. Points corresponding to measurements at the same $\sqrt{s}$ are horizontally shifted for better readability.}
\label{fig:sqrt}
\end{figure}

Figure~\ref{fig:sqrt} presents a summary of the CMS measurements~\cite{Khachatryan:2016mqs,Khachatryan:2016yzq,Khachatryan:2016kzg,CMS:2021vhb,Sirunyan:2018goh} of \stt in \pp collisions at different $\sqrt{s}$ in the $\ell$+jets and dilepton channels, including the one presented in this paper, compared to the NNLO+NNLL
prediction using the NNPDF3.1 NNLO PDF set with $\asmz=0.118$ and $\mt = 172.5\GeV$.
In the inset, the results from this analysis at $\sqrt{s}=5.02\TeV$ are also compared to the predictions from the
MSHT20~\cite{Bailey:2020ooq}, CT18~\cite{Dulat:2015mca}, and ABMP16~\cite{Alekhin:2017kpj} NNLO PDF sets, with the latter using
$\asmz=0.115$ and $\mt = 170.4\GeV$. Theoretical predictions using different PDF sets have comparable
values and uncertainties, once consistent values of \asmz and \mt are associated with the respective PDF set.

The impact of the combined \stt measurement at $\sqrt{s}=5.02\TeV$ on the knowledge of the proton PDFs is tested following the MC methodology of Ref.~\cite{Sirunyan:2017ule}. 
Improvement with respect the baseline fit (i.e., without the inclusion of \ttbar measurements) is observed, verifying the findings of Ref.~\cite{Sirunyan:2017ule} that the $\sigma_\ttbar$ measurements at $\sqrt{s}=5.02\TeV$ are sensitive to the gluon PDF at high Bjorken-$x$ values.

\section{Summary} \label{sec:Summary}

A measurement of the top quark pair production cross section at a center-of-mass energy of 5.02\TeV is presented
for events with one electron and one muon of opposite charge, and at least two jets, using proton-proton collisions collected by the CMS experiment in 2017 and 
corresponding to an integrated luminosity of 302\pbinv. The measured cross section is found to
be $\stt = 60.7 \pm 5.0 \stat \pm 2.8 \syst \pm 1.1 \lum \pb$. A combination with the single lepton + jets measurement, using a data set collected in 2015 at the same center-of-mass energy and corresponding to an integrated luminosity of 27.4\pbinv, is performed. A measurement of $63.0 \pm 4.1 \stat \pm 3.0\, \text{(syst+lumi)}\pb$ is obtained, in agreement with the prediction from the standard model of $66.8^{+2.9}_{-3.1}\pb$.

\begin{acknowledgments}
  We congratulate our colleagues in the CERN accelerator departments for the excellent performance of the LHC and thank the technical and administrative staffs at CERN and at other CMS institutes for their contributions to the success of the CMS effort. In addition, we gratefully acknowledge the computing  centers and personnel of the Worldwide LHC Computing Grid and other  centers for delivering so effectively the computing infrastructure essential to our analyses. Finally, we acknowledge the enduring support for the construction and operation of the LHC, the CMS detector, and the supporting computing infrastructure provided by the following funding agencies: BMBWF and FWF (Austria); FNRS and FWO (Belgium); CNPq, CAPES, FAPERJ, FAPERGS, and FAPESP (Brazil); MES and BNSF (Bulgaria); CERN; CAS, MoST, and NSFC (China); MINCIENCIAS (Colombia); MSES and CSF (Croatia); RIF (Cyprus); SENESCYT (Ecuador); MoER, ERC PUT and ERDF (Estonia); Academy of Finland, MEC, and HIP (Finland); CEA and CNRS/IN2P3 (France); BMBF, DFG, and HGF (Germany); GSRI (Greece); NKFIA (Hungary); DAE and DST (India); IPM (Iran); SFI (Ireland); INFN (Italy); MSIP and NRF (Republic of Korea); MES (Latvia); LAS (Lithuania); MOE and UM (Malaysia); BUAP, CINVESTAV, CONACYT, LNS, SEP, and UASLP-FAI (Mexico); MOS (Montenegro); MBIE (New Zealand); PAEC (Pakistan); MSHE and NSC (Poland); FCT (Portugal); JINR (Dubna); MON, RosAtom, RAS, RFBR, and NRC KI (Russia); MESTD (Serbia); MCIN/AEI and PCTI (Spain); MOSTR (Sri Lanka); Swiss Funding Agencies (Switzerland); MST (Taipei); ThEPCenter, IPST, STAR, and NSTDA (Thailand); TUBITAK and TAEK (Turkey); NASU (Ukraine); STFC (United Kingdom); DOE and NSF (USA).
  
  \hyphenation{Rachada-pisek} Individuals have received support from the Marie-Curie program and the European Research Council and Horizon 2020 Grant, contract Nos.\ 675440, 724704, 752730, 758316, 765710, 824093, 884104, and COST Action CA16108 (European Union); the Leventis Foundation; the Alfred P.\ Sloan Foundation; the Alexander von Humboldt Foundation; the Belgian Federal Science Policy Office; the Fonds pour la Formation \`a la Recherche dans l'Industrie et dans l'Agriculture (FRIA-Belgium); the Agentschap voor Innovatie door Wetenschap en Technologie (IWT-Belgium); the F.R.S.-FNRS and FWO (Belgium) under the ``Excellence of Science -- EOS" -- be.h project n.\ 30820817; the Beijing Municipal Science \& Technology Commission, No. Z191100007219010; the Ministry of Education, Youth and Sports (MEYS) of the Czech Republic; the Deutsche Forschungsgemeinschaft (DFG), under Germany's Excellence Strategy -- EXC 2121 ``Quantum Universe" -- 390833306, and under project number 400140256 - GRK2497; the Lend\"ulet (``Momentum") Prog ram and the J\'anos Bolyai Research Scholarship of the Hungarian Academy of Sciences, the New National Excellence Program \'UNKP, the NKFIA research grants 123842, 123959, 124845, 124850, 125105, 128713, 128786, and 129058 (Hungary); the Council of Science and Industrial Research, India; the Latvian Council of Science; the Ministry of Science and Higher Education and the National Science Center, contracts Opus 2014/15/B/ST2/03998 and 2015/19/B/ST2/02861 (Poland); the Funda\c{c}\~ao para a Ci\^encia e a Tecnologia, grant CEECIND/01334/2018 (Portugal); the National Priorities Research Program by Qatar National Research Fund; the Ministry of Science and Higher Education, projects no. 14.W03.31.0026 and no. FSWW-2020-0008, and the Russian Foundation for Basic Research, project No.19-42-703014 (Russia); MCIN/AEI/10.13039/501100011033, ERDF ``a way of making Europe", and the Programa Estatal de Fomento de la Investigaci{\'o}n Cient{\'i}fica y T{\'e}cnica de Excelencia Mar\'{\i}a de Maeztu, grant MDM-2017-0765 and Programa Severo Ochoa del Principado de Asturias (Spain); the Stavros Niarchos Foundation (Greece); the Rachadapisek Sompot Fund for Postdoctoral Fellowship, Chulalongkorn University and the Chulalongkorn Academic into Its 2nd Century Project Advancement Project (Thailand); the Kavli Foundation; the Nvidia Corporation; the SuperMicro Corporation; the Welch Foundation, contract C-1845; and the Weston Havens Foundation (USA).
\end{acknowledgments}

\bibliography{auto_generated}

\providecommand{\href}[2]{#2}\begingroup\raggedright\begin{thebibliography}{10}%
\makeatletter
\providecommand{\hrefCMSnoop }[0]{\@secondoftwo}%
\makeatother
\providecommand{\doi}{\texttt{doi:}\begingroup \urlstyle{tt}\Url}

\bibitem{Sirunyan:2017azo}
\hrefCMSnoop {}{{CMS Collaboration}, ``{Measurement of double-differential
  cross sections for top quark pair production in \pp collisions at $\sqrt{s} =
  8\TeV$ and impact on parton distribution functions}'',} \textit{ Eur. Phys.
  J. C} \textbf{ 77} (2017) 459,
  \href{http://dx.doi.org/10.1140/epjc/s10052-017-4984-5}{\doi{10.1140/epjc/s10052-017-4984-5}},
\href{http://www.arXiv.org/abs/1703.01630}{\texttt{arXiv:1703.01630}}.

\bibitem{Sirunyan:2018goh}
\hrefCMSnoop {}{{CMS Collaboration}, ``{Measurement of the \ttbar production
  cross section, the top quark mass, and the strong coupling constant using
  dilepton events in \pp collisions at $\sqrt{s}=13\TeV$}'',} \textit{ Eur.
  Phys. J. C} \textbf{ 79} (2019) 368,
  \href{http://dx.doi.org/10.1140/epjc/s10052-019-6863-8}{\doi{10.1140/epjc/s10052-019-6863-8}},
\href{http://www.arXiv.org/abs/1812.10505}{\texttt{arXiv:1812.10505}}.

\bibitem{Sirunyan:2017ule}
\hrefCMSnoop {}{{CMS Collaboration}, ``{Measurement of the inclusive \ttbar
  cross section in \pp collisions at $\sqrt{s}=5.02\TeV$ using final states
  with at least one charged lepton}'',} \textit{ JHEP} \textbf{ 03} (2018) 115,
  \href{http://dx.doi.org/10.1007/JHEP03(2018)115}{\doi{10.1007/JHEP03(2018)115}},
\href{http://www.arXiv.org/abs/1711.03143}{\texttt{arXiv:1711.03143}}.

\bibitem{ATLAStt12}
\hrefCMSnoop {}{{ATLAS Collaboration}, ``{Measurement of the top quark pair
  production cross-section with ATLAS in the single lepton channel}'',}
  \textit{ Phys. Lett. B} \textbf{ 711} (2012) 244,
  \href{http://dx.doi.org/10.1016/j.physletb.2012.03.083}{\doi{10.1016/j.physletb.2012.03.083}},
\href{http://www.arXiv.org/abs/1201.1889}{\texttt{arXiv:1201.1889}}.

\bibitem{Aad:2014kva}
\hrefCMSnoop {}{{ATLAS Collaboration}, ``{Measurement of the \ttbar production
  cross-section using $\Pe{}\Pgm$ events with \PQb{}-tagged jets in \pp
  collisions at $\sqrt{s} = 7$ and 8\TeV with the ATLAS detector}'',} \textit{
  Eur. Phys. J. C} \textbf{ 74} (2014) 3109,
  \href{http://dx.doi.org/10.1140/epjc/s10052-016-4501-2}{\doi{10.1140/epjc/s10052-016-4501-2}},
  \href{http://www.arXiv.org/abs/1406.5375}{\texttt{arXiv:1406.5375}}.
  [Addendum: Eur.Phys.J.C 76, 642 (2016)].

\bibitem{Khachatryan:2016mqs}
\hrefCMSnoop {}{{CMS Collaboration}, ``{Measurement of the \ttbar production
  cross section in the $\Pe{}\Pgm$ channel in proton-proton collisions at
  $\sqrt{s} = 7$ and 8\TeV}'',} \textit{ JHEP} \textbf{ 08} (2016) 029,
  \href{http://dx.doi.org/10.1007/JHEP08(2016)029}{\doi{10.1007/JHEP08(2016)029}},
\href{http://www.arXiv.org/abs/1603.02303}{\texttt{arXiv:1603.02303}}.

\bibitem{Khachatryan:2016yzq}
\hrefCMSnoop {}{{CMS Collaboration}, ``{Measurements of the \ttbar production
  cross section in lepton+jets final states in \pp collisions at 8\TeV and
  ratio of 8 to 7\TeV cross sections}'',} \textit{ Eur. Phys. J. C} \textbf{
  77} (2017) 15,
  \href{http://dx.doi.org/10.1140/epjc/s10052-016-4504-z}{\doi{10.1140/epjc/s10052-016-4504-z}},
\href{http://www.arXiv.org/abs/1602.09024}{\texttt{arXiv:1602.09024}}.

\bibitem{ATLAS:2019hau}
\hrefCMSnoop {}{{ATLAS Collaboration}, ``{Measurement of the \ttbar production
  cross-section and lepton differential distributions in $e\mu$ dilepton events
  from \pp collisions at $\sqrt{s}=13\TeV$ with the ATLAS detector}'',}
  \textit{ Eur. Phys. J. C} \textbf{ 80} (2020) 528,
  \href{http://dx.doi.org/10.1140/epjc/s10052-020-7907-9}{\doi{10.1140/epjc/s10052-020-7907-9}},
  \href{http://www.arXiv.org/abs/1910.08819}{\texttt{arXiv:1910.08819}}.

\bibitem{Khachatryan:2015uqb}
\hrefCMSnoop {}{{CMS Collaboration}, ``{Measurement of the top quark pair
  production cross section in proton-proton collisions at
  $\sqrt{s}=13\TeV$}'',} \textit{ Phys. Rev. Lett.} \textbf{ 116} (2016)
  052002,
  \href{http://dx.doi.org/10.1103/PhysRevLett.116.052002}{\doi{10.1103/PhysRevLett.116.052002}},
\href{http://www.arXiv.org/abs/1510.05302}{\texttt{arXiv:1510.05302}}.

\bibitem{Khachatryan:2016kzg}
\hrefCMSnoop {}{{CMS Collaboration}, ``{Measurement of the \ttbar production
  cross section using events in the $\Pe{}\Pgm$ final state in \pp collisions
  at $\sqrt{s}=13\TeV$}'',} \textit{ Eur. Phys. J. C} \textbf{ 77} (2017) 172,
  \href{http://dx.doi.org/10.1140/epjc/s10052-017-4718-8}{\doi{10.1140/epjc/s10052-017-4718-8}},
\href{http://www.arXiv.org/abs/1611.04040}{\texttt{arXiv:1611.04040}}.

\bibitem{CMS:2021vhb}
\hrefCMSnoop {}{{CMS Collaboration}, ``{Measurement of differential \ttbar
  production cross sections in the full kinematic range using lepton+jets
  events from proton-proton collisions at $\sqrt{s}=13\TeV$}'',} \textit{ Phys.
  Rev. D} \textbf{ 104} (2021) 092013,
  \href{http://dx.doi.org/10.1103/PhysRevD.104.092013}{\doi{10.1103/PhysRevD.104.092013}},
  \href{http://www.arXiv.org/abs/2108.02803}{\texttt{arXiv:2108.02803}}.

\bibitem{Aaij:2018imy}
\hrefCMSnoop {}{{LHCb Collaboration}, ``{Measurement of forward top pair
  production in the dilepton channel in \pp collisions at
  $\sqrt{s}=13\TeV$}'',} \textit{ JHEP} \textbf{ 08} (2018) 174,
  \href{http://dx.doi.org/10.1007/JHEP08(2018)174}{\doi{10.1007/JHEP08(2018)174}},
  \href{http://www.arXiv.org/abs/1803.05188}{\texttt{arXiv:1803.05188}}.

\bibitem{Sirunyan:2017xku}
\hrefCMSnoop {}{{CMS Collaboration}, ``Observation of top quark production in
  proton-nucleus collisions'',} \textit{ Phys. Rev. Lett.} \textbf{ 119} (2017)
  242001,
  \href{http://dx.doi.org/10.1103/PhysRevLett.119.242001}{\doi{10.1103/PhysRevLett.119.242001}},
\href{http://www.arXiv.org/abs/1709.07411}{\texttt{arXiv:1709.07411}}.

\bibitem{Sirunyan:2020kov}
\hrefCMSnoop {}{{CMS Collaboration}, ``{Evidence for top quark production in
  nucleus-nucleus collisions}'',} \textit{ Phys. Rev. Lett.} \textbf{ 125}
  (2020) 222001,
  \href{http://dx.doi.org/10.1103/PhysRevLett.125.222001}{\doi{10.1103/PhysRevLett.125.222001}},
  \href{http://www.arXiv.org/abs/2006.11110}{\texttt{arXiv:2006.11110}}.

\bibitem{hepdata}
\href {https://www.hepdata.net/record/102986}{``Hepdata record for this
  analysis'',} 2021.
\newblock \url {https://www.hepdata.net/record/102986}.

\bibitem{Chatrchyan:2008zzk}
\hrefCMSnoop {}{{CMS Collaboration}, ``The {CMS} experiment at the {CERN}
  {LHC}'',} \textit{ JINST} \textbf{ 3} (2008) S08004,
  \href{http://dx.doi.org/10.1088/1748-0221/3/08/S08004}{\doi{10.1088/1748-0221/3/08/S08004}}.

\bibitem{Agostinelli:2002hh}
\hrefCMSnoop {}{{GEANT4} Collaboration, ``{\GEANTfour}---a simulation
  toolkit'',} \textit{ Nucl. Instrum. Meth. A} \textbf{ 506} (2003) 250,
\href{http://dx.doi.org/10.1016/S0168-9002(03)01368-8}{\doi{10.1016/S0168-9002(03)01368-8}}.

\bibitem{Frixione:2007vw}
\hrefCMSnoop {}{S.~Frixione, P.~Nason, and C.~Oleari, ``{Matching NLO QCD
  computations with parton shower simulations: the \POWHEG method}'',} \textit{
  JHEP} \textbf{ 11} (2007) 070,
  \href{http://dx.doi.org/10.1088/1126-6708/2007/11/070}{\doi{10.1088/1126-6708/2007/11/070}},
\href{http://www.arXiv.org/abs/0709.2092}{\texttt{arXiv:0709.2092}}.

\bibitem{Alioli:2010xd}
\hrefCMSnoop {}{S.~Alioli, P.~Nason, C.~Oleari, and E.~Re, ``{A general
  framework for implementing NLO calculations in shower Monte Carlo programs:
  the \POWHEG BOX}'',} \textit{ JHEP} \textbf{ 06} (2010) 043,
  \href{http://dx.doi.org/10.1007/JHEP06(2010)043}{\doi{10.1007/JHEP06(2010)043}},
\href{http://www.arXiv.org/abs/1002.2581}{\texttt{arXiv:1002.2581}}.

\bibitem{Frixione:2007nw}
\hrefCMSnoop {}{S.~Frixione, P.~Nason, and G.~Ridolfi, ``{A positive-weight
  next-to-leading-order Monte Carlo for heavy flavor hadroproduction}'',}
  \textit{ JHEP} \textbf{ 09} (2007) 126,
  \href{http://dx.doi.org/10.1088/1126-6708/2007/09/126}{\doi{10.1088/1126-6708/2007/09/126}},
\href{http://www.arXiv.org/abs/0707.3088}{\texttt{arXiv:0707.3088}}.

\bibitem{Sjostrand:2014zea}
T.~Sj{\"o}strand\hrefCMSnoop {}{ {et~al.}, ``{An introduction to
  \PYTHIA8.2}'',} \textit{ Comput. Phys. Commun.} \textbf{ 191} (2015) 159,
  \href{http://dx.doi.org/10.1016/j.cpc.2015.01.024}{\doi{10.1016/j.cpc.2015.01.024}},
\href{http://www.arXiv.org/abs/1410.3012}{\texttt{arXiv:1410.3012}}.

\bibitem{Sirunyan:2019dfx}
\hrefCMSnoop {}{{CMS Collaboration}, ``{Extraction and validation of a new set
  of CMS \PYTHIA8 tunes from underlying-event measurements}'',} \textit{ Eur.
  Phys. J. C} \textbf{ 80} (2020) 4,
  \href{http://dx.doi.org/10.1140/epjc/s10052-019-7499-4}{\doi{10.1140/epjc/s10052-019-7499-4}},
\href{http://www.arXiv.org/abs/1903.12179}{\texttt{arXiv:1903.12179}}.

\bibitem{Ball:2017nwa}
\hrefCMSnoop {}{{NNPDF} Collaboration, ``{Parton distributions from
  high-precision collider data}'',} \textit{ Eur. Phys. J. C} \textbf{ 77}
  (2017) 663,
  \href{http://dx.doi.org/10.1140/epjc/s10052-017-5199-5}{\doi{10.1140/epjc/s10052-017-5199-5}},
\href{http://www.arXiv.org/abs/1706.00428}{\texttt{arXiv:1706.00428}}.

\bibitem{Alwall:2014hca}
J.~Alwall\hrefCMSnoop {}{ {et~al.}, ``{The automated computation of tree-level
  and next-to-leading order differential cross sections, and their matching to
  parton shower simulations}'',} \textit{ JHEP} \textbf{ 07} (2014) 079,
  \href{http://dx.doi.org/10.1007/JHEP07(2014)079}{\doi{10.1007/JHEP07(2014)079}},
\href{http://www.arXiv.org/abs/1405.0301}{\texttt{arXiv:1405.0301}}.

\bibitem{Frederix:2012ps}
\hrefCMSnoop {}{R.~Frederix and S.~Frixione, ``{Merging meets matching in
  \MCATNLO}'',} \textit{ JHEP} \textbf{ 12} (2012) 061,
  \href{http://dx.doi.org/10.1007/JHEP12(2012)061}{\doi{10.1007/JHEP12(2012)061}},
\href{http://www.arXiv.org/abs/1209.6215}{\texttt{arXiv:1209.6215}}.

\bibitem{Czakon:2011xx}
\hrefCMSnoop {}{M.~Czakon and A.~Mitov, ``{\TOPpp: a program for the
  calculation of the top-pair cross-section at hadron colliders}'',} \textit{
  Comput. Phys. Commun.} \textbf{ 185} (2014) 2930,
  \href{http://dx.doi.org/10.1016/j.cpc.2014.06.021}{\doi{10.1016/j.cpc.2014.06.021}},
\href{http://www.arXiv.org/abs/1112.5675}{\texttt{arXiv:1112.5675}}.

\bibitem{Czakon:2013goa}
\hrefCMSnoop {}{M.~Czakon, P.~Fiedler, and A.~Mitov, ``{Total top quark pair
  production cross section at hadron colliders through O($\alpS^4$)}'',}
  \textit{ Phys. Rev. Lett.} \textbf{ 110} (2013) 252004,
  \href{http://dx.doi.org/10.1103/PhysRevLett.110.252004}{\doi{10.1103/PhysRevLett.110.252004}},
\href{http://www.arXiv.org/abs/1303.6254}{\texttt{arXiv:1303.6254}}.

\bibitem{Kidonakis:2015nna}
\hrefCMSnoop {}{N.~Kidonakis, ``{Theoretical results for electroweak-boson and
  single-top production}'',} \textit{ PoS} \textbf{ DIS2015} (2015) 170,
\href{http://www.arXiv.org/abs/1506.04072}{\texttt{arXiv:1506.04072}}.

\bibitem{Melnikov:2006kv}
\hrefCMSnoop {}{K.~Melnikov and F.~Petriello, ``{Electroweak gauge boson
  production at hadron colliders through $O(\alpS^2)$}'',} \textit{ Phys. Rev.
  D} \textbf{ 74} (2006) 114017,
  \href{http://dx.doi.org/10.1103/PhysRevD.74.114017}{\doi{10.1103/PhysRevD.74.114017}},
\href{http://www.arXiv.org/abs/hep-ph/0609070}{\texttt{arXiv:hep-ph/0609070}}.

\bibitem{PDG2020}
\hrefCMSnoop {}{{Particle Data Group} Collaboration, ``{Review of Particle
  Physics}'',} \textit{ PTEP} \textbf{ 2020} (2020) 083C01,
  \href{http://dx.doi.org/10.1093/ptep/ptaa104}{\doi{10.1093/ptep/ptaa104}}.

\bibitem{Khachatryan:2016bia}
\hrefCMSnoop {}{{CMS Collaboration}, ``{The CMS trigger system}'',} \textit{
  JINST} \textbf{ 12} (2017) P01020,
  \href{http://dx.doi.org/10.1088/1748-0221/12/01/P01020}{\doi{10.1088/1748-0221/12/01/P01020}},
\href{http://www.arXiv.org/abs/1609.02366}{\texttt{arXiv:1609.02366}}.

\bibitem{Sirunyan:2020zal}
\hrefCMSnoop {}{{CMS Collaboration}, ``{Performance of the CMS Level-1 trigger
  in proton-proton collisions at $\sqrt{s}=13\TeV$}'',} \textit{ JINST}
  \textbf{ 15} (2020) P10017,
  \href{http://dx.doi.org/10.1088/1748-0221/15/10/P10017}{\doi{10.1088/1748-0221/15/10/P10017}},
  \href{http://www.arXiv.org/abs/2006.10165}{\texttt{arXiv:2006.10165}}.

\bibitem{Cacciari:2008gp}
\hrefCMSnoop {}{M.~Cacciari, G.~P. Salam, and G.~Soyez, ``The anti-\kt jet
  clustering algorithm'',} \textit{ JHEP} \textbf{ 04} (2008) 063,
  \href{http://dx.doi.org/10.1088/1126-6708/2008/04/063}{\doi{10.1088/1126-6708/2008/04/063}},
  \href{http://www.arXiv.org/abs/0802.1189}{\texttt{arXiv:0802.1189}}.

\bibitem{Cacciari:2011ma}
\hrefCMSnoop {}{M.~Cacciari, G.~P. Salam, and G.~Soyez, ``{\FASTJET user
  manual}'',} \textit{ Eur. Phys. J. C} \textbf{ 72} (2012) 1896,
  \href{http://dx.doi.org/10.1140/epjc/s10052-012-1896-2}{\doi{10.1140/epjc/s10052-012-1896-2}},
\href{http://www.arXiv.org/abs/1111.6097}{\texttt{arXiv:1111.6097}}.

\bibitem{CMS-PRF-14-001}
\hrefCMSnoop {}{{CMS Collaboration}, ``Particle-flow reconstruction and global
  event description with the {CMS} detector'',} \textit{ JINST} \textbf{ 12}
  (2017) P10003,
  \href{http://dx.doi.org/10.1088/1748-0221/12/10/P10003}{\doi{10.1088/1748-0221/12/10/P10003}},
\href{http://www.arXiv.org/abs/1706.04965}{\texttt{arXiv:1706.04965}}.

\bibitem{Khachatryan:2016kdb}
\hrefCMSnoop {}{{CMS Collaboration}, ``{Jet energy scale and resolution in the
  CMS experiment in \pp collisions at 8\TeV}'',} \textit{ JINST} \textbf{ 12}
  (2017) P02014,
  \href{http://dx.doi.org/10.1088/1748-0221/12/02/P02014}{\doi{10.1088/1748-0221/12/02/P02014}},
  \href{http://www.arXiv.org/abs/1607.03663}{\texttt{arXiv:1607.03663}}.

\bibitem{CMS:2017wyc}
\href {https://cds.cern.ch/record/2256875}{{CMS Collaboration}, ``{Jet
  algorithms performance in 13\TeV data}'',} CMS Physics Analysis Summary
  CMS-PAS-JME-16-003, 2017.

\bibitem{CMS:2020uim}
\hrefCMSnoop {}{{CMS Collaboration}, ``{Electron and photon reconstruction and
  identification with the CMS experiment at the CERN LHC}'',} \textit{ JINST}
  \textbf{ 16} (2021) P05014,
  \href{http://dx.doi.org/10.1088/1748-0221/16/05/P05014}{\doi{10.1088/1748-0221/16/05/P05014}},
  \href{http://www.arXiv.org/abs/2012.06888}{\texttt{arXiv:2012.06888}}.

\bibitem{Sirunyan:2018fpa}
\hrefCMSnoop {}{{CMS Collaboration}, ``{Performance of the CMS muon detector
  and muon reconstruction with proton-proton collisions at
  $\sqrt{s}=13\TeV$}'',} \textit{ JINST} \textbf{ 13} (2018) P06015,
  \href{http://dx.doi.org/10.1088/1748-0221/13/06/P06015}{\doi{10.1088/1748-0221/13/06/P06015}},
  \href{http://www.arXiv.org/abs/1804.04528}{\texttt{arXiv:1804.04528}}.

\bibitem{Sirunyan:2020icl}
\hrefCMSnoop {}{{CMS Collaboration}, ``{Measurement of the Higgs boson
  production rate in association with top quarks in final states with
  electrons, muons, and hadronically decaying tau leptons at
  $\sqrt{s}=13\TeV$}'',} \textit{ Eur. Phys. J. C} \textbf{ 81} (2021) 378,
  \href{http://dx.doi.org/10.1140/epjc/s10052-021-09014-x}{\doi{10.1140/epjc/s10052-021-09014-x}},
  \href{http://www.arXiv.org/abs/2011.03652}{\texttt{arXiv:2011.03652}}.

\bibitem{Sirunyan:2017ezt}
\hrefCMSnoop {}{{CMS Collaboration}, ``{Identification of heavy-flavour jets
  with the CMS detector in \pp collisions at 13\TeV}'',} \textit{ JINST}
  \textbf{ 13} (2018) P05011,
  \href{http://dx.doi.org/10.1088/1748-0221/13/05/P05011}{\doi{10.1088/1748-0221/13/05/P05011}},
  \href{http://www.arXiv.org/abs/1712.07158}{\texttt{arXiv:1712.07158}}.

\bibitem{Butterworth:2015oua}
\hrefCMSnoop {}{J.~Butterworth {et~al.}, ``{PDF4LHC recommendations for LHC Run
  II}'',} \textit{ J. Phys. G} \textbf{ 43} (2016) 023001,
  \href{http://dx.doi.org/10.1088/0954-3899/43/2/023001}{\doi{10.1088/0954-3899/43/2/023001}},
  \href{http://www.arXiv.org/abs/1510.03865}{\texttt{arXiv:1510.03865}}.

\bibitem{Sirunyan:2018nqx}
\hrefCMSnoop {}{{CMS Collaboration}, ``{Measurement of the inelastic
  proton-proton cross section at $\sqrt{s}=13\TeV$}'',} \textit{ JHEP} \textbf{
  07} (2018) 161,
  \href{http://dx.doi.org/10.1007/JHEP07(2018)161}{\doi{10.1007/JHEP07(2018)161}},
\href{http://www.arXiv.org/abs/1802.02613}{\texttt{arXiv:1802.02613}}.

\bibitem{bib:LUM-19-001}
\href {https://cds.cern.ch/record/2765655}{{CMS Collaboration}, ``{Luminosity
  calibration for the \pp reference run at $\sqrt{s}=5.02\TeV$ in 2017}'',} CMS
  Physics Analysis Summary CMS-PAS-LUM-19-001, 2019.

\bibitem{Lyons:1988rp}
\hrefCMSnoop {}{L.~Lyons, D.~Gibaut, and P.~Clifford, ``{How to combine
  correlated estimates of a single physical quantity}'',} \textit{ Nucl.
  Instrum. Meth. A} \textbf{ 270} (1988) 110,
  \href{http://dx.doi.org/10.1016/0168-9002(88)90018-6}{\doi{10.1016/0168-9002(88)90018-6}}.

\bibitem{Valassi:2013bga}
\hrefCMSnoop {}{A.~Valassi and R.~Chierici, ``{Information and treatment of
  unknown correlations in the combination of measurements using the BLUE
  method}'',} \textit{ Eur. Phys. J. C} \textbf{ 74} (2014) 2717,
  \href{http://dx.doi.org/10.1140/epjc/s10052-014-2717-6}{\doi{10.1140/epjc/s10052-014-2717-6}},
  \href{http://www.arXiv.org/abs/1307.4003}{\texttt{arXiv:1307.4003}}.

\bibitem{Lista:2014qia}
\hrefCMSnoop {}{L.~Lista, ``{The bias of the unbiased estimator: a study of the
  iterative application of the BLUE method}'',} \textit{ Nucl. Instrum. Meth.
  A} \textbf{ 764} (2014) 82,
  \href{http://dx.doi.org/10.1016/j.nima.2014.07.021}{\doi{10.1016/j.nima.2014.07.021}},
  \href{http://www.arXiv.org/abs/1405.3425}{\texttt{arXiv:1405.3425}}.
  [Erratum: \DOI{10.1016/j.nima.2014.11.054}].

\bibitem{Bailey:2020ooq}
S.~Bailey\hrefCMSnoop {}{ {et~al.}, ``{Parton distributions from LHC, HERA,
  Tevatron and fixed target data: MSHT20 PDFs}'',} \textit{ Eur. Phys. J. C}
  \textbf{ 81} (2021) 341,
  \href{http://dx.doi.org/10.1140/epjc/s10052-021-09057-0}{\doi{10.1140/epjc/s10052-021-09057-0}},
  \href{http://www.arXiv.org/abs/2012.04684}{\texttt{arXiv:2012.04684}}.

\bibitem{Dulat:2015mca}
S.~Dulat\hrefCMSnoop {}{ {et~al.}, ``{New parton distribution functions from a
  global analysis of quantum chromodynamics}'',} \textit{ Phys. Rev. D}
  \textbf{ 93} (2016) 033006,
  \href{http://dx.doi.org/10.1103/PhysRevD.93.033006}{\doi{10.1103/PhysRevD.93.033006}},
\href{http://www.arXiv.org/abs/1506.07443}{\texttt{arXiv:1506.07443}}.

\bibitem{Alekhin:2017kpj}
\hrefCMSnoop {}{S.~Alekhin, J.~Bl{\"u}mlein, S.~Moch, and R.~Placakyte,
  ``{Parton distribution functions, \alpS, and heavy-quark masses for LHC Run
  II}'',} \textit{ Phys. Rev. D} \textbf{ 96} (2017) 014011,
  \href{http://dx.doi.org/10.1103/PhysRevD.96.014011}{\doi{10.1103/PhysRevD.96.014011}},
\href{http://www.arXiv.org/abs/1701.05838}{\texttt{arXiv:1701.05838}}.

\end{thebibliography}\endgroup

\cleardoublepage \appendix\section{The CMS Collaboration \label{app:collab}}\begin{sloppypar}\hyphenpenalty=5000\widowpenalty=500\clubpenalty=5000\cmsinstitute{Yerevan~Physics~Institute, Yerevan, Armenia}
A.~Tumasyan
\cmsinstitute{Institut~f\"{u}r~Hochenergiephysik, Vienna, Austria}
W.~Adam\cmsorcid{0000-0001-9099-4341}, J.W.~Andrejkovic, T.~Bergauer\cmsorcid{0000-0002-5786-0293}, S.~Chatterjee\cmsorcid{0000-0003-2660-0349}, M.~Dragicevic\cmsorcid{0000-0003-1967-6783}, A.~Escalante~Del~Valle\cmsorcid{0000-0002-9702-6359}, R.~Fr\"{u}hwirth\cmsAuthorMark{1}, M.~Jeitler\cmsAuthorMark{1}\cmsorcid{0000-0002-5141-9560}, N.~Krammer, L.~Lechner\cmsorcid{0000-0002-3065-1141}, D.~Liko, I.~Mikulec, P.~Paulitsch, F.M.~Pitters, J.~Schieck\cmsAuthorMark{1}\cmsorcid{0000-0002-1058-8093}, R.~Sch\"{o}fbeck\cmsorcid{0000-0002-2332-8784}, M.~Spanring\cmsorcid{0000-0001-6328-7887}, S.~Templ\cmsorcid{0000-0003-3137-5692}, W.~Waltenberger\cmsorcid{0000-0002-6215-7228}, C.-E.~Wulz\cmsAuthorMark{1}\cmsorcid{0000-0001-9226-5812}
% ...
\vskip\cmsinstskip
% ...
1:~Also at TU Wien, Wien, Austria\\
% ...
\end{sloppypar}
\end{document}